\documentclass[12pt]{article}
\begin{document}
\title{\bf Cylindrically Symmetric, Static, Perfect-Fluid Solutions of
Einstein's Field Equations}

\author{M. Sharif \thanks{E-mail: sharifm@hepth.hanyang.ac.kr,
Department of Mathematics, Punjab University,
Quid-e-Azam Campus, Lahore-54590, Pakistan}\\
Department of Physics, Hanyang University, Seoul 133-791.}

\date{}
\maketitle

Perfect-fluid, static, cylindrically symmetric solutions of
Einstein's field equations are obtained for the equations of state
$\rho+3p=0$ and $\rho=p$. In the former case, the density and the
pressure turn out to be constant while in the later case, they
depend on the radial parameter $r$. Evan's solution corresponding
to the equation of state $\rho=p$ is included in the solutions
discussed here.

 \vspace{0.5cm}

Solutions of Einstein's field equations for given physical
situations are not easy to obtain on account of their non-linearity.
Many exact solutions have been obtained [1] by using a simple form
of the stress-energy tensor and assuming some symmetries. Here, we
take a perfect fluid with the equations of state and assume
staticity and cylindrical symmetry. There have been pervious
discussions of a similar nature by Evan [2], Bronnikov [3], Latelier
and Tobensky [4], and Kramer [5]. They used various equations of
state that could be written in the form $\rho=\gamma p$ for specific
positive values of $\gamma$, as well as energy conservation. Our
procedure is to use the general form of the static, cylindrically
symmetric metric
\begin{equation}
ds^2=e^{\nu(r)}dt^2-dr^2-e^{\lambda(r)}d\theta^2-e^{\mu(r)}dz^2
\end{equation}
with the stress-energy tensor
\begin{equation}
T_{ab}=(\rho+p)\dot{x}_a\dot{x}_b-pg_{ab}~(a,b=0,1,2,3),
\end{equation}
where the four-vector velocity $\dot{x}_a$, is given by $e^{\nu/2}
\delta^{0}_{a}$. Einstein's equations then become
\begin{eqnarray}
-8\pi\rho&=&e^{-\lambda/2}(e^{-\lambda/2})''
+e^{-\mu/2}(e^{-\mu/2})''+\frac {\lambda'\mu'}{4},\\
8\pi p&=&e^{-\mu 2}(e^{\mu/2})''+e^{-\nu/2}(e^{\nu/2})''
+\frac{\mu'\nu'}{4},\\
8\pi p&=&e^{-\nu/2}(e^{\nu/2})''+e ^{-\lambda/2}
(e^{\lambda/2})''+\frac{\lambda'\nu'}{4},\\
8\pi p&=&\frac{1}{4}(\lambda'\mu'+\mu\nu+\lambda\nu),
\end{eqnarray}
where the dash represents the derivative relative to the radial
parameter. We assume further that one of the three arbitrary
functions becomes a constant and deduce the equation of state to
which the solution corresponds. Eqs. (4)-(6) can be rewritten as the
set of 3 linearly dependent equations
\begin{eqnarray}
e^{-\mu/2}(e^{\mu/2})''-e ^{-\lambda/2}(e^{\lambda/2})''+\frac {\
\nu'(\mu'- \lambda')}{4}&=&0,\\
e^{-\mu/2}(e^{\mu/2})''e ^{-\nu/2}(e^{\nu/2})''-\frac {\
\lambda'(\mu'+\nu')}{4}&=&0,\\
e^{-\nu/2}(e^{\nu/2})''+e ^{-\lambda/2}(e^{\lambda/2})''-\frac {\
\mu'(\lambda'+\nu')}{4}&=&0,
\end{eqnarray}
and Eq.(3) can be used to obtain an equation for $\rho$. We consider
only the solutions of these equations for the cases (A) $\nu=0$, (B)
$\lambda=0$ and (C) $\mu=0$. \\
\textbf{Case(A):} Eqs.(7)-(9), in this case reduce, respectively, to
\begin{eqnarray}
e^{-\mu/2}(e^{\mu/2})''-e ^{-\lambda/2}(e^{\lambda/2})''&=&0,\\
(e^{\mu/2})'&=&k_1e^{\lambda/2},\\
(e^{\lambda/2})'&=&k_2e^{\mu/2},
\end{eqnarray}
where $k_{i}~(i=1,2,3...)$ are constants of integration. With Eqs.
(11)-(12), Eq.(10) reduces to
\begin{equation}
k_{1}e^{\lambda}-k_{2}e^{\mu}=k_{3},
\end{equation}
which serves as a constraint on the solution of Eqs.(11)-(12). The
solution of Eqs.(11)-(12) then appears as
\begin{eqnarray}
e^{\mu/2}&=&k_5 \cosh(\sqrt{k_4}r+k_6),\quad k_4>0,\nonumber\\
&=&-k_5\cos(\sqrt{-k_4}r+k_6),\quad k_4>0,\nonumber\\
&=&k_5r+k_6,\quad k_4=0,\quad k_1\neq0,\quad k_2=0,\nonumber\\
&=&k_5,\quad k_4=0=k_1,
\end{eqnarray}
and correspondingly
\begin{eqnarray}
e^{\lambda/2}&=&k_5\sqrt{\frac{k_2}{k1}}\sinh(\sqrt{k_4}r
+k_6),\quad k_4>0,\nonumber\\
&=&-k_5\sqrt{-\frac{k_2}{k_1}}\sin(\sqrt{-k_4}r+k_6),\quad k_4<0,\nonumber\\
&=&k_2k_5r+k_6,\quad k_4=0,k_1=0,\quad k_2\neq0,\nonumber\\
&=&k_7,\quad k_4=0=k_ 2,
\end{eqnarray}
where $k_4=-k_1k_2$ and $k_3=-k^2_5$ for $k_4<0$ or $k_4>0$ or
$k_4=0,~k_2\neq0;~k_3=k_1k_7$ for $k_4=0,~k_1\neq0$, and
$k_1=0=k_2,~k_3=0$ for $k_4=0$. It is worth noting at this stage
that for each of these solutions
$e^{-\lambda/2}(e^{\lambda/2})''=e^{-\mu/2}(e^{\mu/2})''=
\frac{\lambda' \mu'}{4}=-k_4$. Thus, from Eqs.(3)-(6),
$\rho=\frac{3}{8 \pi}k_4$ and $p=-\frac{k_4}{8 \pi}$, showing that
$\rho+3 p=0$. The solution with $k_4<0$ has a negative energy
density, and that with $k_4>0$ has a positive energy density, where
as with $k_4=0$, the solutions are trivial. Since the energy density
is constant, its graph will be a straight line. Also, the
relationship between the energy density and the pressure gives a
straight line.

\textbf{Case B:} In this case, Eqs.(7)-(9) are reduced to
\begin{eqnarray}
\{e^{\nu/2}(e^{\mu/2})'\}' &=&0,\\
e^{-{\mu}{2}}(e^{{\mu}{2}})''+e^{-\nu/2}(e^{\nu/2})''&=&0,\\
\{e^{-\mu/2}(e^{\nu/2})'\}'&=&0,
\end{eqnarray}
Thus, from Eqs.(16) and (18), we get
\begin{eqnarray}
(e^{\mu/2})'&=&k_1e^{-\nu/2},\\
(e^{\nu/2})'&=&k_2e^{\mu/2},
\end{eqnarray}
and due to Eqs.(19)-(20), Eq.(17) is identically satisfied for $\mu$
and $\nu$ uniquely determined by
\begin{equation}
e^{\nu/2}(e^{\nu/2})''=k_1k_2
\end{equation}
and one of the Eqs.(19)-(20). The solution of Eq.(21) is determined
implicitly by the integral
\begin{equation}
\int\frac{d(e^{\nu/2})}{\sqrt{k_3-k_1k_2\frac{v}{2}}}=r+k_{4,}
\end{equation}
which reduces to
\begin{equation}
\rm{erf}(\frac{y}{\sqrt{k_1k_2}})=-k_4r+k_{5},
\end{equation}
where $y=\sqrt{k_3-k_1k_2\frac{\nu}{2}}$ and
$k_4=\frac{1}{2}k_1k_2\exp(-\frac{k_3}{k_1k_2})$. The convergence of
the integral is ensured if $k_1k_2>0$, for whivh
\begin{equation}
\rho=\frac{1}{8\pi}k_1k_2e^{-\nu}=p.
\end{equation}
We see that the energy density is not a constant, rather, it depends
on the value of $\nu$, which is a function of $r$. Thus, in this
case, the geometry of $\rho$ may not be a straight line, but will
depend on the value of $\nu$. However, as $\rho$ and $p$ are equal,
the graph between them will be a straight line.

Case (C) is now trivial and can be soled by replacing $\nu$ and
$\lambda$ in case (B); this case then yields the Evan's solution
[2]. In this case, the geometrical situation will be the same as for
case (B) as we have $\rho=p$.

It is of special interest to note that the case of $\rho+3p=0$ gives
a constant pressure and density. (Remember that pure radiation
corresponds to $\rho-3p=0$). Further, the case of $\rho=p$ has been
considered earlier [2] and appears as case (C) discussed here.

\vspace{0.5cm}

{\bf Acknowledgments}

\vspace{0.5cm}

The author would like to thank Prof. Chul H. Lee for his hospitality
at the Department of Physics and Korea Science and Engineering
Foundation (KOSEF) for his postdoc fellowship at Hanyang University,
Seoul, Korea. This work was also supported by grant No.
1999-2-11200-003-5 from the Basic Research Program of KOSEF.

\vspace{0.5cm}

{\bf \Large References}

\begin{description}

\item{[1]} D. Kramer, H. Stephani, M. MacCallum and E. Herlt:
\emph{Exact Solutions of Einstein's Field Equations} (Cambridge
University Press, Cambridge, 1980).

\item{[2]} A.B. Evan: J. Phys. \textbf{A10}(1977)1303.

\item{[3]} K.A. Bronnikov: J. Phys. \textbf{A12}(1979)201.

\item{[4]} P.S. Laterlier and R.R. Tabensky: Nuovo Cimento \textbf{B28}(1975)407.

\item{[5]} D. Kramer: Class. Quantum Grav. \textbf{5}(1988)393.
\end{description}
\end{document}